\documentclass[jkps,twocolumn,showpacs,showkeys,superscriptaddress]{revtex4}
\usepackage[pdftex]{graphicx}
\usepackage{xcolor}
\usepackage{amssymb}
\usepackage{amsmath, amsfonts}
\usepackage{bm}
\usepackage{soul}
\usepackage{ulem}

\begin{document}

\title[]{\textit{In-situ} scanning tunneling microscopy observation of thickness-dependent air-sensitive layered materials and heterodevices}

\author{Hyoung Kug Kim}
\author{Dowook Kim}
\author{Dong Guk Lee}
\affiliation{Department of Physics, Pohang University of Science and Technology (POSTECH), Pohang, 37673, Korea}
\author{Eun-Su Ahn}
\affiliation{Department of Physics, Pohang University of Science and Technology (POSTECH), Pohang, 37673, Korea}
\affiliation{Center for Artificial Low Dimensional Electronic Systems, Institute for Basic Science (IBS), Pohang 37673, Korea}
\author{Hyeon-Woo Jeong}
\author{Gil-Ho Lee}
\affiliation{Department of Physics, Pohang University of Science and Technology (POSTECH), Pohang, 37673, Korea}
\author{Jun Sung Kim}
\affiliation{Department of Physics, Pohang University of Science and Technology (POSTECH), Pohang, 37673, Korea}
\affiliation{Center for Artificial Low Dimensional Electronic Systems, Institute for Basic Science (IBS), Pohang 37673, Korea}
\author{Tae-Hwan Kim}
\email[Electronic mail: ]{taehwan@postech.ac.kr}
\affiliation{Department of Physics, Pohang University of Science and Technology (POSTECH), Pohang, 37673, Korea}

\date{\today}

\begin{abstract}
Quasi-two-dimensional (Quasi-2D) van der Waals (vdW) materials can be mechanically or chemically exfoliated down to monolayer because of their strong intralayer bonding and the weak interlayer vdW interaction.
Thanks to this unique property, one can often find exotic thickness-dependent electronic properties from these quasi-2D vdW materials,
which can lead to band gap opening, emerging superconductivity, or enhanced charge density waves with decreasing thickness.
Surface-sensitive scanning tunneling microscopy (STM) can provide direct observation of structural and electronic characteristics of such layered materials with atomic precision in real space.
However, it is very challenging to preserve the intrinsic surfaces of air-sensitive quasi-2D materials between preparation and measurement.
In addition, vdW 2D crystals after exfoliation are extremely hard to explore with a typical STM setup due to their small size ($<10~\mu$m).
Here, we present a straightforward method compatible with any STM setup having optical access:
(1) exfoliating and/or stacking layered materials in a glove box,
(2) transferring them to an ultra-high vacuum STM chamber using a suitcase without exposure to air, 
and (3) navigating surface to locate exfoliated vdW 2D flakes with different thicknesses.
We successfully demonstrated that the clean surfaces of the air-sensitive Fe$_3$GeTe$_2$ can be effectively protected from unwanted oxidation during transfer. 
Furthermore, our method provides a simple but useful way to access a specific tiny stack of layered materials without any \textit{ex-situ} fabrication processes for STM navigation.
Our experimental improvement will open up a new way to investigate air-sensitive layered vdW materials with various thicknesses via surface-sensitive techniques including STM.
\end{abstract}



\pagenumbering{roman} 

\maketitle

\clearpage
\pagenumbering{arabic}

\section{Introduction}
Quasi-two-dimensional (Quasi-2D) materials, for example, graphene or transition metal dichalcogenides (TMDs), feature strong intralayer bondings and weak interlayer van der Waals (vdW) interaction.
The weak vdW coupling between the interlayers enables layered vdW crystals to be exfoliated down to monolayer, which can exhibit exotic electronic states from massless Dirac fermion~\cite{Geim2009} to a dimer-induced insulator~\cite{Hwang2022}.
The dimensionality-driven transition of electronic orders in exfoliated vdW materials also includes a tunable electronic band gap~\cite{Terrones2013} or enhanced superconductivity~\cite{Rhodes2021} and charge density waves~\cite{Feng2018, Xi2015}.
To unveil such remarkable electronic characteristics, macroscopic or spatially averaged measurements have generally been developed, including electronic/magnetic transport~\cite{Lin2015a, Cao2018a, Sharpe2019} and photoluminescence~\cite{Lin2015a, Seyler2019}.
On the other hand, microscopic local probe measurements, such as scanning tunneling microscopy (STM), have rarely been employed even though they are essential to investigate both surface-related and local electronic states with atomic spatial resolution, for example, the correlation between moir\'e superlattices with superconductivity or spatially varying band gaps~\cite{Fan2015, Choi2019}.

To explore the intrinsic properties with atomic precision in real space using surface-sensitive techniques including STM, the surface cleanliness of ultra-thin materials is highly required.
In this sense, inert quasi-2D vdW materials such as graphene~\cite{Yankowitz2012, Woods2014} and TMD semiconductors~\cite{Hill2016, Shih2017} are widely investigated because the materials are rarely oxidized during the sample preparation in air.
In contrast to inert vdW materials, several techniques have been developed to prevent the rapid degradation in air-sensitive 2D vdW materials: deposition of capping layers~\cite{Liu2015, Liu2015a} before exposure to air and \textit{in-situ} exfoliation of layered materials in ultra-high vacuum (UHV)~\cite{Pasztor2017}.
However, the former method requires the thermal desorption of the capping layer, which often causes contamination or damage on the pristine surface of layered material.
However, the latter requires a time-consuming search for exfoliated flakes since the typical optical setup of an UHV STM chamber can hardly identify tiny flakes ($<10~\mu$m) under STM~\cite{Pasztor2017}.
Instead of the time-consuming navigation, one can fabricate guiding patterns toward a target flake~\cite{Zhao2015} but cannot avoid unwanted contamination due to such an \textit{ex-situ} fabrication process.

Here, we introduce a straightforward method to successfully preserve the clean surface of exfoliated quasi-2D vdW crystals, including highly oxygen-reactive TMDs, via a combination of a standard glove box and a portable suitcase.
This method provides direct access to ultra-thin and tiny layered vdW materials within a relatively short time by comparing the optical microscope (OM) images taken in the glove box with optical telescope (OT) images in an STM setup.
We can observe the surface of air-sensitive Fe$_3$GeTe$_2$ (FGT) without any exposure to air between the sample preparation and measurements.
Furthermore, we demonstrate that our method can allow us to perform thickness-dependent research on air-sensitive quasi-2D materials.
Additionally, stacked heterostructures such as graphene on hexagonal boron nitride (hBN) can be investigated by our method.
Our findings would offer an alternative way to investigate air-sensitive layered vdW materials with different thicknesses and diverse twisted-heterojunction devices via surface-sensitive techniques.
For example, our method can be directly applied to chemically vulnerable black phosphorus, which exhibits interesting thickness-dependent band gaps~\cite{Qiao2014, Xia2019}.

\section{Experimental details}
To prevent the oxidation of 2D vdW materials,
the exfoliation and transfer processes are performed using a standard glove box filled with pure Ar gas (H$_2$O~$<$~0.05~ppm, O$_2$~$<$~0.05~ppm) 
and a portable suitcase (0.5$\times$0.3$\times$0.2~m$^3$).
To demonstrate the effectiveness of our method, we chose air-sensitive FGT~\cite{Kim2018, Kim2019} as follows.

\begin{figure}[ht]
\includegraphics[width = \linewidth]{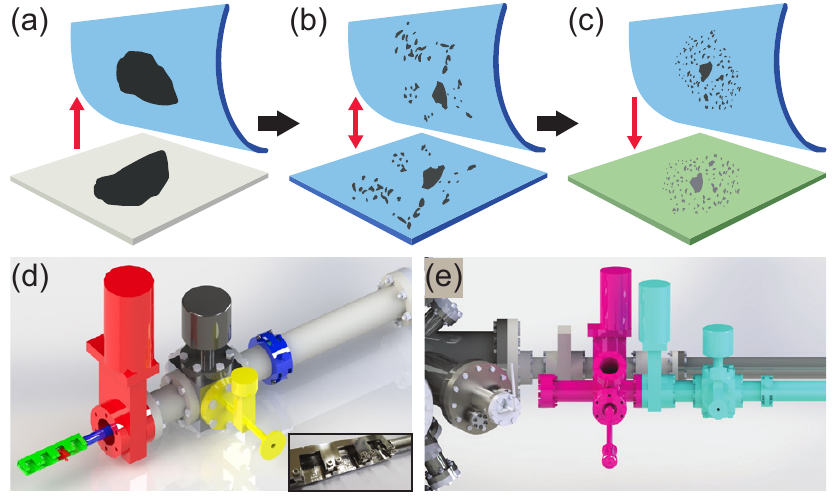}
\caption{Exfoliation of 2D materials and portable suitcase. 
(a)--(c) Schematic drawing of exfoliating thin 2D  materials in an Ar-filled glove box.
See more details in the text.
(d) 3D CAD drawing of a compact-sized portable suitcase consisting of four receptacles for sample storage (green), a gate valve (red), a bellows-sealed valve (yellow), a circular handle (gray), and a transfer arm (blue).
The inset shows a zoom-in photograph of the sample receptacles.
(e) 3D CAD drawing of the suitcase (cyan) and the junction chamber (magenta) attached to the load-lock chamber (light gray).
}
\label{fig1}
\end{figure}

In an Ar-filled glove box, we carefully cleaved and transferred a mother FGT crystal on a standard tape (gray) onto a residue-free clean-room adhesive tape (Nitto BT-150E-CM, blue) [Fig.~\ref{fig1}(a)].
The flake was then repeatedly cleaved with a new residue-free tape to obtain thin enough flakes [Fig.~\ref{fig1}(b)], typically ranging from a few to a hundred nanometers.
The obtained thin FGT flakes were finally transferred onto a graphene substrate that was epitaxially grown on a silicon carbide (G/SiC, green) single crystal (Cree Inc., W4NPF0X-0200) [Fig.~\ref{fig1}(c)].
The substrate with thin FGT flakes was mounted on a sample holder for STM measurements.
Once a sample holder was ready for transfer, we took several OM images to easily locate thin flakes for STM measurements (see below). 

We used a home-built portable suitcase to carry FGT flakes from a glove box to an STM chamber.
To do so, the suitcase is small enough to pass through a glove box entrance.
As shown in Fig.~\ref{fig1}(d), we could assemble a portable suitcase consisting of a 6-way cube, a bellows-sealed valve (Swagelok, SS-4BG), a gate valve, four receptacles for sample storage, a circular handle for carrying, and a transfer arm.
Note that all components of the suitcase are UHV-compatible and bakeable up to 400~K, leading to a base pressure below $5 \times 10^{-10}$~torr.
Before introducing the suitcase into a glove box, we first pumped the suitcase containing empty sample holders.
After being introduced into a glove box, the suitcase was carefully ventilated with pure Ar gas in the glove box through the bellows-sealed valve.
Subsequently, the sample holders with FGT flakes were rigidly stored inside the suitcase [Fig.~\ref{fig1}(d)].
Next, with all valves closed, the Ar-filled suitcase was moved out of the glove box and then attached to the load-lock chamber next to our STM chamber [Fig.~\ref{fig1}(e)].
Lastly, we transferred the sample holders from the suitcase to the STM chamber through the load-lock chamber after Ar gas in the suitcase was pumped out.
In this way, we could prepare clean and thin FGT flakes without any exposure to air between exfoliation and STM measurements.

\section{Results and Discussion}
In general, mechanical exfoliation reduces not only the thickness of flakes but also their lateral size.
We can obtain exfoliated FGT flakes with a typical lateral size of $\approx 10~\mu$m and a thickness of 10--100~nm.
The increase in the density of exfoliated flakes is beneficial to STM measurements.
To do so, flakes need to be exfoliated too many times, which often leads to poor surface quality due to tape residue. 
Instead, we should locate a target FGT flake to perform STM measurements of such FGT flakes with low density.

\begin{figure}[ht]
\includegraphics[width = \linewidth]{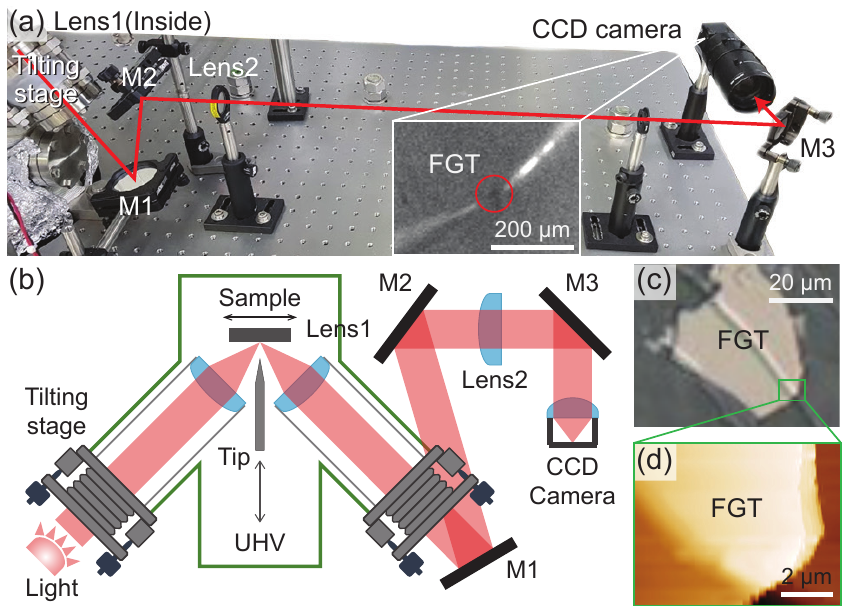}
\caption{Experimental setup and how to locate thin flakes.
(a) Photograph of the experimental setup showing an optical navigation system next to the STM chamber.
The inset shows a CCD image of the tip apex and its mirror image on the target FGT flake (indicated by a red circle).
(b) Schematic drawing of the optical navigation system consisting of two convex lenses with fixed focal lengths, mirrors, and a CCD camera with a $\times10$ zoom lens.
(c) Zoom-in OM image and (d) STM topographic image of the target FGT flake.
The green square indicates the corresponding location in both cases.
}
\label{fig2}
\end{figure}

Figure~2(a) illustrates a simple optical setup to navigate the surface under STM.
We use convex lenses with fixed focal lengths ($f_1$ = 150~mm and $f_2$ = 300~mm) and one zoom lens with variable magnification from $\times$1 to $\times$10 before a $\times$2 extender to provide the total magnification ranging from $\times$4 to $\times$40.
The field of view can reach as large as $3.0\times2.6$~mm$^2$ at $\times40$ magnification.
We installed distance-adjustable tilting stages holding the lenses in a vacuum as shown in Fig.~2(b).
Thanks to the tilting stages, we can perform a fine adjustment during optical alignment without breaking the vacuum.
When cooling an STM head, the suspension springs are slightly shortened due to thermal contraction. 
However, such a change can be compensated by the tilting stages, too.
For optical access, the inner and outer radiation shields surrounding the STM head have opening holes with a diameter of 5~mm.
Note that the openings increase the STM base temperature by only $\sim$ 3~K. 
Once locating a flake, we can close the openings entirely to reach the base temperature for STM measurements.

To find and investigate a specific flake, we first focus on a large and thick flake [Fig.~2(c)], which can be observed through a CCD image of the optical setup in our STM [see the inset of Fig.~2(a)].
Once locating the target flake, we bring the STM tip on it by moving the STM sample stage with respect to the STM tip.
Then, the automated STM approach can successfully bring the STM tip to the flake surface.
Finally, we can verify the precise location of the flake in the scan window [Figs.~2(c) and 2(d)]. 
After finding the precise location, we can move the STM tip toward thinner flakes next to the large and thick flake using the previously taken OM images in a globe box.

\begin{figure}[ht]
\includegraphics[width = \linewidth]{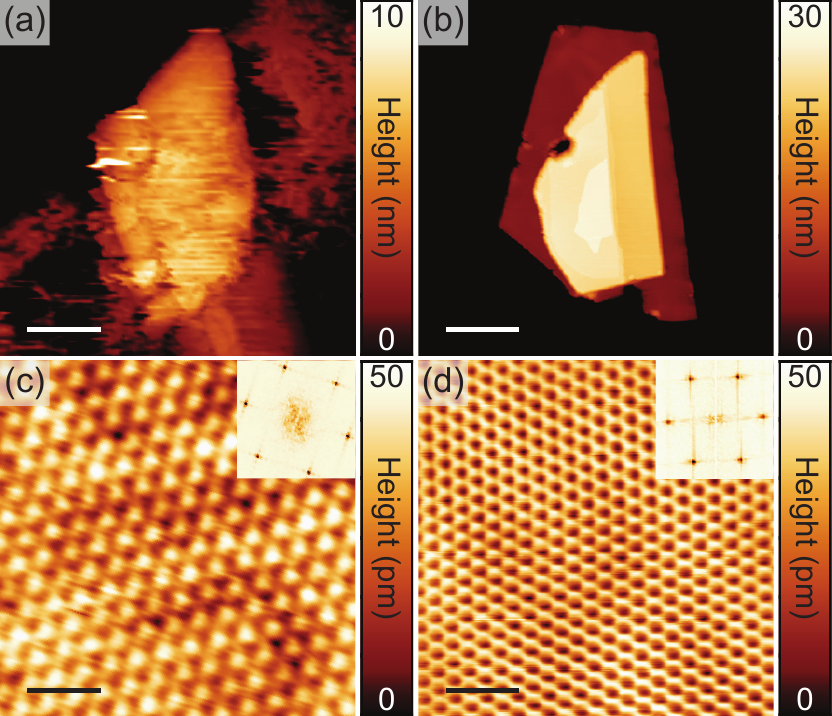}
\caption{Effectiveness of a suitcase during sample transfer.
(a) [(b)] STM image of a thin air-sensitive FGT flake without (with) using the suitcase during the sample transfer from the glove box to the load-lock chamber. 
While a short ($<$ 5~min) exposure to air results in oxidizing FGT flakes [(a)],
the Ar-filled suitcase effectively prevents them from oxidation during the transfer,
leading to atomically flat and clean surfaces [(b)].
(c) Atom-resolved STM image of the topmost FGT flake in (b) reveals the vacancy-free hexagonal Te lattice ($V_{\text{b}}$ = 0.5~V, $I_{\text{t}}$ = 6.0~nA).
(d) Atom-resolved STM image of graphene near the FGT flake showing a honeycomb lattice ($V_{\text{b}}$ = 1.0~V, $I_{\text{t}}$ = 10~pA).
Insets of (c) and (d) show the corresponding fast Fourier transform (FFT) images of (c) and (d), respectively.
The FFT images exhibit the six-fold rotational symmetry of each surface.
Scale bars denote (a) 40~nm, (b) 180~nm, (c) 1~nm, and (d) 1~nm.
}
\label{fig3}
\end{figure}

Before moving further, we would like to discuss the effectiveness of the suitcase during sample transfer.
To demonstrate how effective the use of the suitcase is, we perform STM experiments of air-sensitive FGT flakes transferred without and with using the suitcase between a glove box and an STM chamber.
As clearly seen, the exfoliated FGT flakes in a glove box are heavily oxidized without the suitcase even with quick exposure to air ($<5$~min) [Fig.~\ref{fig3}(a)] 
while the suitcase allows us to observe flat and clean surfaces without any signature of oxidation [Fig.~\ref{fig3}(b)].
This distinct difference strongly indicates that both a glove box and a suitcase are crucial to avoid unwanted oxidation.
Furthermore, we can confirm that the intact FGT flake shows a hexagonal Te atomic lattice in the atom-resolved STM image when the FGT flakes are carried by the suitcase [Fig.~\ref{fig3}(c)].
The STM image exhibits characteristic bright sub-layer Fe vacancy defects ($< 30$~pm in height) just below the defect-free Te topmost layer as shown in bulk FGT cleaved in UHV~\cite{Kim2018},
which suggests that our method is comparable to the \textit{in-situ} UHV exfoliation.

\begin{figure}[ht]
\includegraphics[width = \linewidth]{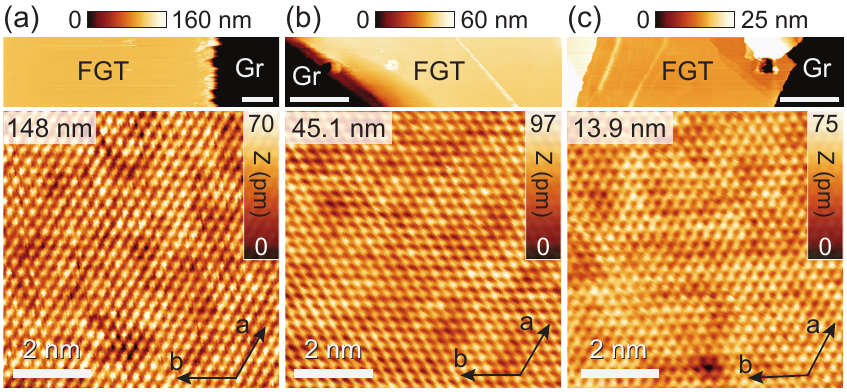}
\caption{Thin FGT flakes with different thicknesses exfoliated from the same mother crystal.
(a--c) Large-scale STM topographic and corresponding atom-resolved images with thicknesses of 148~nm, 45.1~nm, and 13.9~nm, respectively.
Imaging conditions for atom-resolved images were set to $V_{\text{b}}$ = 0.1--0.3~V and $I_{\text{t}}$ = 0.5--8.0~nA.
Scale bars on all large-scale and atom-resolved STM images denote 0.5~$\mu$m and 2.0~nm, respectively. 
}
\label{fig4}
\end{figure}

As discussed above, we can obtain thin FGT flakes of different thicknesses from the same mother crystal and perform thickness-dependent STM measurements.
We would like to emphasize that all flakes were naturally inherited from the mother crystal.
This capability provides a unique opportunity for studying thickness-dependent properties of pristine FGT without worrying about sample inhomogeneity.
Otherwise, we should carefully investigate the sample inhomogeneity in a crystal-by-crystal manner~\cite{May2016} to isolate the thickness effect.

Figure~4 shows three different FGT flakes with the thicknesses of 148~nm, 45~nm, and 14~nm. 
The thickness-dependent STM topography demonstrates that all flakes ranging from 10~nm to 150~nm have a similar atomic structure, which is quite consistent with the previously reported magnetic properties~\cite{Fei2018, Deng2018}.
In this experiment, we could not find thinner flakes than 4~nm, which may exhibit somewhat different structural features being related to the observed 2D Ising ferromagnetism~\cite{Fei2018, Deng2018}.
This successful microscopic thickness-dependent measurement can be applied to other quasi-2D materials that feature sophisticated distinct properties with different thicknesses.

\begin{figure}[ht]
\includegraphics[width = \linewidth]{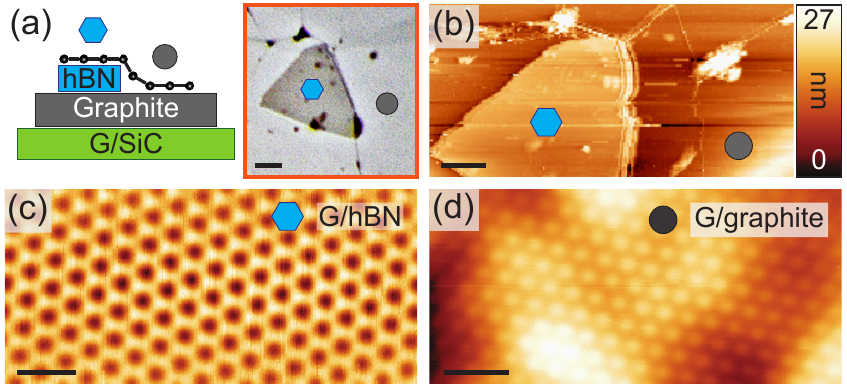}
\caption{Stacked heterostructure of different vdW materials.
(a) Schematic of a stacked heterostructure consisting of graphene (G, represented by small circles and lines), hBN (blue rectangle), and a graphite flake (Gr, by gray rectangle) on G/SiC (by green rectangle).
Blue hexagon (gray circle) indicates the G/hBN (G/Gr) heterostructure.
Inset shows an OM image showing the stacked heterostructure of G/hBN/Gr on G/SiC.
The smaller triangular hBN is sandwiched between (top) graphene and (bottom) graphite.
(b) STM topographic image of the stacked heterostructure of G/hBN/Gr ($V_{\text{b}}$ = $-3.0$~V, $I_{\text{t}}$ = 50~pA).
(c) [(d)] Zoom-in STM image of the moir\'e superlattice of G/hBN (G/Gr).
Scale bars denote (a,b) 2~$\mu$m and (c,d) 3.0~nm. 
}
\label{fig5}
\end{figure}

In addition to FGT flakes, we can apply this method to stacked 2D heterostructures such as a graphene/hBN/graphite assembly (G/hBN/Gr)~\cite{note}, which are impossible to assemble through the simple \textit{in-situ} UHV exfoliation~\cite{Pasztor2017}.
It is also very challenging to directly locate the tiny stacked 2D heterostructure in a typical STM setup.
However, in the same way as described in Fig.~2, the quick positioning of the STM tip on the stacked 2D heterostructure can be successfully conducted.
As shown in the schematic drawing and OM image [Fig.~\ref{fig5}(a)], a graphene monolayer fully covers a triangular hBN flake (indicated by a blue hexagon) on a graphite flake while the graphite flake is partially covered by the graphene layer (by a gray circle).
The graphene monolayer is too transparent to be detected in the OM image, but large-scale STM images reveal the triangular hBN crystal covered with graphene [Fig.~\ref{fig5}(b)].
Furthermore, zoom-in STM images in Figs.~\ref{fig5}(c) and \ref{fig5}(d) exhibit two different moir\'e structures corresponding to vertically stacked G/hBN and G/Gr layers, respectively.
From the measured moir\'e superlattice periodicities (1.4~nm and 1.2~nm),
we can estimate the twisted angles between the top and bottom layers (G/hBN and G/Gr)~\cite{Yankowitz2012},
which are 20.4$^{\circ}$ and 11.9$^{\circ}$, respectively.
These estimated angles are well consistent with those measured between the well-defined crystalline edges of each layer in the OM images.
Thus, we can explore any stacked 2D heterostructure with our experimental approach.

\section{Summary}
We present a straightforward method to prepare air-sensitive thin 2D flakes for surface-sensitive measurements using a glove box and a portable suitcase.
In addition, we employ a simple optical setup to bring the STM tip  toward a specific flake for thickness dependent measurements.
The feasibility and effectiveness of our method is well demonstrated by exfoliating and imaging air-sensitive FGT flakes with different thicknesses.
Thanks to our method, all FGT flakes exhibit intact and defect-free topmost Te layers, otherwise being oxidized even with short exposure to air.
Furthermore, we can explore a stacked heterostructure device with STM,
which is typically inaccessible without \textit{ex-situ} guiding patterns in the STM study.
Our experimental improvement will pave a new route to the diverse, precise, and microscopic thickness-dependent study of various surface-sensitive layered materials as well as complicated stacked devices.

\begin{acknowledgments}
This work was supported by the National Research Foundation of Korea grant funded by the Korea government (MSIT) (NRF-2017R1A2B4007742, 2021R1A6A1A10042944, 2022R1A2C3009731, and 2022M3H4A1A04074153).
H.-W.J. and G.-H.L. were supported by the MSIT under the Information Technology Research Center support program (IITP-2022-RS-2022-00164799) supervised by the Institute for Information \& Communications Technology Planning \& Evaluation.
\end{acknowledgments}


\begin{thebibliography}{28}
\expandafter\ifx\csname natexlab\endcsname\relax\def\natexlab#1{#1}\fi
\expandafter\ifx\csname bibnamefont\endcsname\relax
  \def\bibnamefont#1{#1}\fi
\expandafter\ifx\csname bibfnamefont\endcsname\relax
  \def\bibfnamefont#1{#1}\fi
\expandafter\ifx\csname citenamefont\endcsname\relax
  \def\citenamefont#1{#1}\fi
\expandafter\ifx\csname url\endcsname\relax
  \def\url#1{\texttt{#1}}\fi
\expandafter\ifx\csname urlprefix\endcsname\relax\def\urlprefix{URL }\fi
\providecommand{\bibinfo}[2]{#2}
\providecommand{\eprint}[2][]{\url{#2}}

\bibitem[{\citenamefont{Geim}(2009)}]{Geim2009}
\bibinfo{author}{\bibfnamefont{A.~K.} \bibnamefont{Geim}},
  \bibinfo{journal}{Science} \textbf{\bibinfo{volume}{324}},
  \bibinfo{pages}{1530} (\bibinfo{year}{2009}).

\bibitem[{\citenamefont{Hwang et~al.}(2022)\citenamefont{Hwang, Kim, Zhang,
  Zhu, Herbig, Kim, Kim, Zhong, Salah, El-Desoky et~al.}}]{Hwang2022}
\bibinfo{author}{\bibfnamefont{J.}~\bibnamefont{Hwang}},
  \bibinfo{author}{\bibfnamefont{K.}~\bibnamefont{Kim}},
  \bibinfo{author}{\bibfnamefont{C.}~\bibnamefont{Zhang}},
  \bibinfo{author}{\bibfnamefont{T.}~\bibnamefont{Zhu}},
  \bibinfo{author}{\bibfnamefont{C.}~\bibnamefont{Herbig}},
  \bibinfo{author}{\bibfnamefont{S.}~\bibnamefont{Kim}},
  \bibinfo{author}{\bibfnamefont{B.}~\bibnamefont{Kim}},
  \bibinfo{author}{\bibfnamefont{Y.}~\bibnamefont{Zhong}},
  \bibinfo{author}{\bibfnamefont{M.}~\bibnamefont{Salah}},
  \bibinfo{author}{\bibfnamefont{M.~M.} \bibnamefont{El-Desoky}},
  \bibnamefont{et~al.}, \bibinfo{journal}{Nat. Commun.}
  \textbf{\bibinfo{volume}{13}}, \bibinfo{pages}{906} (\bibinfo{year}{2022}).

\bibitem[{\citenamefont{Terrones et~al.}(2013)\citenamefont{Terrones,
  L{\'{o}}pez-Ur{\'{i}}as, and Terrones}}]{Terrones2013}
\bibinfo{author}{\bibfnamefont{H.}~\bibnamefont{Terrones}},
  \bibinfo{author}{\bibfnamefont{F.}~\bibnamefont{L{\'{o}}pez-Ur{\'{i}}as}},
  \bibnamefont{and} \bibinfo{author}{\bibfnamefont{M.}~\bibnamefont{Terrones}},
  \bibinfo{journal}{Sci. Rep.} \textbf{\bibinfo{volume}{3}},
  \bibinfo{pages}{1549} (\bibinfo{year}{2013}).

\bibitem[{\citenamefont{Rhodes et~al.}(2021)\citenamefont{Rhodes, Jindal, Yuan,
  Jung, Antony, Wang, Kim, Chiu, Taniguchi, Watanabe et~al.}}]{Rhodes2021}
\bibinfo{author}{\bibfnamefont{D.~A.} \bibnamefont{Rhodes}},
  \bibinfo{author}{\bibfnamefont{A.}~\bibnamefont{Jindal}},
  \bibinfo{author}{\bibfnamefont{N.~F.~Q.} \bibnamefont{Yuan}},
  \bibinfo{author}{\bibfnamefont{Y.}~\bibnamefont{Jung}},
  \bibinfo{author}{\bibfnamefont{A.}~\bibnamefont{Antony}},
  \bibinfo{author}{\bibfnamefont{H.}~\bibnamefont{Wang}},
  \bibinfo{author}{\bibfnamefont{B.}~\bibnamefont{Kim}},
  \bibinfo{author}{\bibfnamefont{Y.-c.} \bibnamefont{Chiu}},
  \bibinfo{author}{\bibfnamefont{T.}~\bibnamefont{Taniguchi}},
  \bibinfo{author}{\bibfnamefont{K.}~\bibnamefont{Watanabe}},
  \bibnamefont{et~al.}, \bibinfo{journal}{Nano Lett.}
  \textbf{\bibinfo{volume}{21}}, \bibinfo{pages}{2505} (\bibinfo{year}{2021}).

\bibitem[{\citenamefont{Feng et~al.}(2018)\citenamefont{Feng, Biswas, Rajan,
  Watson, Mazzola, Clark, Underwood, Marković, McLaren, Hunter
  et~al.}}]{Feng2018}
\bibinfo{author}{\bibfnamefont{J.}~\bibnamefont{Feng}},
  \bibinfo{author}{\bibfnamefont{D.}~\bibnamefont{Biswas}},
  \bibinfo{author}{\bibfnamefont{A.}~\bibnamefont{Rajan}},
  \bibinfo{author}{\bibfnamefont{M.~D.} \bibnamefont{Watson}},
  \bibinfo{author}{\bibfnamefont{F.}~\bibnamefont{Mazzola}},
  \bibinfo{author}{\bibfnamefont{O.~J.} \bibnamefont{Clark}},
  \bibinfo{author}{\bibfnamefont{K.}~\bibnamefont{Underwood}},
  \bibinfo{author}{\bibfnamefont{I.}~\bibnamefont{Marković}},
  \bibinfo{author}{\bibfnamefont{M.}~\bibnamefont{McLaren}},
  \bibinfo{author}{\bibfnamefont{A.}~\bibnamefont{Hunter}},
  \bibnamefont{et~al.}, \bibinfo{journal}{Nano Lett.}
  \textbf{\bibinfo{volume}{18}}, \bibinfo{pages}{4493} (\bibinfo{year}{2018}).

\bibitem[{\citenamefont{Xi et~al.}(2015)\citenamefont{Xi, Zhao, Wang, Berger,
  Forró, Shan, and Mak}}]{Xi2015}
\bibinfo{author}{\bibfnamefont{X.}~\bibnamefont{Xi}},
  \bibinfo{author}{\bibfnamefont{L.}~\bibnamefont{Zhao}},
  \bibinfo{author}{\bibfnamefont{Z.}~\bibnamefont{Wang}},
  \bibinfo{author}{\bibfnamefont{H.}~\bibnamefont{Berger}},
  \bibinfo{author}{\bibfnamefont{L.}~\bibnamefont{Forró}},
  \bibinfo{author}{\bibfnamefont{J.}~\bibnamefont{Shan}}, \bibnamefont{and}
  \bibinfo{author}{\bibfnamefont{K.~F.} \bibnamefont{Mak}},
  \bibinfo{journal}{Nat. Nanotechnol.} \textbf{\bibinfo{volume}{10}},
  \bibinfo{pages}{765} (\bibinfo{year}{2015}).

\bibitem[{\citenamefont{Lin et~al.}(2015)\citenamefont{Lin, Ghosh, Addou, Lu,
  Eichfeld, Zhu, Li, Peng, Kim, Li et~al.}}]{Lin2015a}
\bibinfo{author}{\bibfnamefont{Y.-C.} \bibnamefont{Lin}},
  \bibinfo{author}{\bibfnamefont{R.~K.} \bibnamefont{Ghosh}},
  \bibinfo{author}{\bibfnamefont{R.}~\bibnamefont{Addou}},
  \bibinfo{author}{\bibfnamefont{N.}~\bibnamefont{Lu}},
  \bibinfo{author}{\bibfnamefont{S.~M.} \bibnamefont{Eichfeld}},
  \bibinfo{author}{\bibfnamefont{H.}~\bibnamefont{Zhu}},
  \bibinfo{author}{\bibfnamefont{M.-Y.} \bibnamefont{Li}},
  \bibinfo{author}{\bibfnamefont{X.}~\bibnamefont{Peng}},
  \bibinfo{author}{\bibfnamefont{M.~J.} \bibnamefont{Kim}},
  \bibinfo{author}{\bibfnamefont{L.-J.} \bibnamefont{Li}},
  \bibnamefont{et~al.}, \bibinfo{journal}{Nat. Commun.}
  \textbf{\bibinfo{volume}{6}}, \bibinfo{pages}{7311} (\bibinfo{year}{2015}).

\bibitem[{\citenamefont{Cao et~al.}(2018)\citenamefont{Cao, Fatemi, Fang,
  Watanabe, Taniguchi, Kaxiras, and Jarillo-Herrero}}]{Cao2018a}
\bibinfo{author}{\bibfnamefont{Y.}~\bibnamefont{Cao}},
  \bibinfo{author}{\bibfnamefont{V.}~\bibnamefont{Fatemi}},
  \bibinfo{author}{\bibfnamefont{S.}~\bibnamefont{Fang}},
  \bibinfo{author}{\bibfnamefont{K.}~\bibnamefont{Watanabe}},
  \bibinfo{author}{\bibfnamefont{T.}~\bibnamefont{Taniguchi}},
  \bibinfo{author}{\bibfnamefont{E.}~\bibnamefont{Kaxiras}}, \bibnamefont{and}
  \bibinfo{author}{\bibfnamefont{P.}~\bibnamefont{Jarillo-Herrero}},
  \bibinfo{journal}{Nature} \textbf{\bibinfo{volume}{556}}, \bibinfo{pages}{43}
  (\bibinfo{year}{2018}).

\bibitem[{\citenamefont{Sharpe et~al.}(2019)\citenamefont{Sharpe, Fox, Barnard,
  Finney, Watanabe, Taniguchi, Kastner, and Goldhaber-Gordon}}]{Sharpe2019}
\bibinfo{author}{\bibfnamefont{A.~L.} \bibnamefont{Sharpe}},
  \bibinfo{author}{\bibfnamefont{E.~J.} \bibnamefont{Fox}},
  \bibinfo{author}{\bibfnamefont{A.~W.} \bibnamefont{Barnard}},
  \bibinfo{author}{\bibfnamefont{J.}~\bibnamefont{Finney}},
  \bibinfo{author}{\bibfnamefont{K.}~\bibnamefont{Watanabe}},
  \bibinfo{author}{\bibfnamefont{T.}~\bibnamefont{Taniguchi}},
  \bibinfo{author}{\bibfnamefont{M.~A.} \bibnamefont{Kastner}},
  \bibnamefont{and}
  \bibinfo{author}{\bibfnamefont{D.}~\bibnamefont{Goldhaber-Gordon}},
  \bibinfo{journal}{Science} \textbf{\bibinfo{volume}{365}},
  \bibinfo{pages}{605} (\bibinfo{year}{2019}).

\bibitem[{\citenamefont{Seyler et~al.}(2019)\citenamefont{Seyler, Rivera, Yu,
  Wilson, Ray, Mandrus, Yan, Yao, and Xu}}]{Seyler2019}
\bibinfo{author}{\bibfnamefont{K.~L.} \bibnamefont{Seyler}},
  \bibinfo{author}{\bibfnamefont{P.}~\bibnamefont{Rivera}},
  \bibinfo{author}{\bibfnamefont{H.}~\bibnamefont{Yu}},
  \bibinfo{author}{\bibfnamefont{N.~P.} \bibnamefont{Wilson}},
  \bibinfo{author}{\bibfnamefont{E.~L.} \bibnamefont{Ray}},
  \bibinfo{author}{\bibfnamefont{D.~G.} \bibnamefont{Mandrus}},
  \bibinfo{author}{\bibfnamefont{J.}~\bibnamefont{Yan}},
  \bibinfo{author}{\bibfnamefont{W.}~\bibnamefont{Yao}}, \bibnamefont{and}
  \bibinfo{author}{\bibfnamefont{X.}~\bibnamefont{Xu}},
  \bibinfo{journal}{Nature} \textbf{\bibinfo{volume}{567}}, \bibinfo{pages}{66}
  (\bibinfo{year}{2019}).

\bibitem[{\citenamefont{Fan et~al.}(2015)\citenamefont{Fan, Zhang, Liu, Yan,
  Ren, Xia, Chen, Xu, Ye, Jiao et~al.}}]{Fan2015}
\bibinfo{author}{\bibfnamefont{Q.}~\bibnamefont{Fan}},
  \bibinfo{author}{\bibfnamefont{W.~H.} \bibnamefont{Zhang}},
  \bibinfo{author}{\bibfnamefont{X.}~\bibnamefont{Liu}},
  \bibinfo{author}{\bibfnamefont{Y.~J.} \bibnamefont{Yan}},
  \bibinfo{author}{\bibfnamefont{M.~Q.} \bibnamefont{Ren}},
  \bibinfo{author}{\bibfnamefont{M.}~\bibnamefont{Xia}},
  \bibinfo{author}{\bibfnamefont{H.~Y.} \bibnamefont{Chen}},
  \bibinfo{author}{\bibfnamefont{D.~F.} \bibnamefont{Xu}},
  \bibinfo{author}{\bibfnamefont{Z.~R.} \bibnamefont{Ye}},
  \bibinfo{author}{\bibfnamefont{W.~H.} \bibnamefont{Jiao}},
  \bibnamefont{et~al.}, \bibinfo{journal}{Phys. Rev. B}
  \textbf{\bibinfo{volume}{91}}, \bibinfo{pages}{104506}
  (\bibinfo{year}{2015}).

\bibitem[{\citenamefont{Choi et~al.}(2019)\citenamefont{Choi, Kemmer, Peng,
  Thomson, Arora, Polski, Zhang, Ren, Alicea, Refael et~al.}}]{Choi2019}
\bibinfo{author}{\bibfnamefont{Y.}~\bibnamefont{Choi}},
  \bibinfo{author}{\bibfnamefont{J.}~\bibnamefont{Kemmer}},
  \bibinfo{author}{\bibfnamefont{Y.}~\bibnamefont{Peng}},
  \bibinfo{author}{\bibfnamefont{A.}~\bibnamefont{Thomson}},
  \bibinfo{author}{\bibfnamefont{H.}~\bibnamefont{Arora}},
  \bibinfo{author}{\bibfnamefont{R.}~\bibnamefont{Polski}},
  \bibinfo{author}{\bibfnamefont{Y.}~\bibnamefont{Zhang}},
  \bibinfo{author}{\bibfnamefont{H.}~\bibnamefont{Ren}},
  \bibinfo{author}{\bibfnamefont{J.}~\bibnamefont{Alicea}},
  \bibinfo{author}{\bibfnamefont{G.}~\bibnamefont{Refael}},
  \bibnamefont{et~al.}, \bibinfo{journal}{Nat. Phys.}
  \textbf{\bibinfo{volume}{15}}, \bibinfo{pages}{1174} (\bibinfo{year}{2019}).

\bibitem[{\citenamefont{Yankowitz et~al.}(2012)\citenamefont{Yankowitz, Xue,
  Cormode, Sanchez-Yamagishi, Watanabe, Taniguchi, Jarillo-Herrero, Jacquod,
  and LeRoy}}]{Yankowitz2012}
\bibinfo{author}{\bibfnamefont{M.}~\bibnamefont{Yankowitz}},
  \bibinfo{author}{\bibfnamefont{J.}~\bibnamefont{Xue}},
  \bibinfo{author}{\bibfnamefont{D.}~\bibnamefont{Cormode}},
  \bibinfo{author}{\bibfnamefont{J.~D.} \bibnamefont{Sanchez-Yamagishi}},
  \bibinfo{author}{\bibfnamefont{K.}~\bibnamefont{Watanabe}},
  \bibinfo{author}{\bibfnamefont{T.}~\bibnamefont{Taniguchi}},
  \bibinfo{author}{\bibfnamefont{P.}~\bibnamefont{Jarillo-Herrero}},
  \bibinfo{author}{\bibfnamefont{P.}~\bibnamefont{Jacquod}}, \bibnamefont{and}
  \bibinfo{author}{\bibfnamefont{B.~J.} \bibnamefont{LeRoy}},
  \bibinfo{journal}{Nat. Phys.} \textbf{\bibinfo{volume}{8}},
  \bibinfo{pages}{382} (\bibinfo{year}{2012}).

\bibitem[{\citenamefont{Woods et~al.}(2014)\citenamefont{Woods, Britnell,
  Eckmann, Ma, Lu, Guo, Lin, Yu, Cao, Gorbachev et~al.}}]{Woods2014}
\bibinfo{author}{\bibfnamefont{C.~R.} \bibnamefont{Woods}},
  \bibinfo{author}{\bibfnamefont{L.}~\bibnamefont{Britnell}},
  \bibinfo{author}{\bibfnamefont{A.}~\bibnamefont{Eckmann}},
  \bibinfo{author}{\bibfnamefont{R.~S.} \bibnamefont{Ma}},
  \bibinfo{author}{\bibfnamefont{J.~C.} \bibnamefont{Lu}},
  \bibinfo{author}{\bibfnamefont{H.~M.} \bibnamefont{Guo}},
  \bibinfo{author}{\bibfnamefont{X.}~\bibnamefont{Lin}},
  \bibinfo{author}{\bibfnamefont{G.~L.} \bibnamefont{Yu}},
  \bibinfo{author}{\bibfnamefont{Y.}~\bibnamefont{Cao}},
  \bibinfo{author}{\bibfnamefont{R.~V.} \bibnamefont{Gorbachev}},
  \bibnamefont{et~al.}, \bibinfo{journal}{Nat. Phys.}
  \textbf{\bibinfo{volume}{10}}, \bibinfo{pages}{451} (\bibinfo{year}{2014}).

\bibitem[{\citenamefont{Hill et~al.}(2016)\citenamefont{Hill, Rigosi, Rim,
  Flynn, and Heinz}}]{Hill2016}
\bibinfo{author}{\bibfnamefont{H.~M.} \bibnamefont{Hill}},
  \bibinfo{author}{\bibfnamefont{A.~F.} \bibnamefont{Rigosi}},
  \bibinfo{author}{\bibfnamefont{K.~T.} \bibnamefont{Rim}},
  \bibinfo{author}{\bibfnamefont{G.~W.} \bibnamefont{Flynn}}, \bibnamefont{and}
  \bibinfo{author}{\bibfnamefont{T.~F.} \bibnamefont{Heinz}},
  \bibinfo{journal}{Nano Lett.} \textbf{\bibinfo{volume}{16}},
  \bibinfo{pages}{4831} (\bibinfo{year}{2016}).

\bibitem[{\citenamefont{Shih et~al.}(2017)\citenamefont{Shih, Li, Ren, Zhang,
  Jin, Chou, Chuu, and Li}}]{Shih2017}
\bibinfo{author}{\bibfnamefont{C.-K.} \bibnamefont{Shih}},
  \bibinfo{author}{\bibfnamefont{L.-J.} \bibnamefont{Li}},
  \bibinfo{author}{\bibfnamefont{X.}~\bibnamefont{Ren}},
  \bibinfo{author}{\bibfnamefont{C.}~\bibnamefont{Zhang}},
  \bibinfo{author}{\bibfnamefont{C.}~\bibnamefont{Jin}},
  \bibinfo{author}{\bibfnamefont{M.-Y.} \bibnamefont{Chou}},
  \bibinfo{author}{\bibfnamefont{C.-P.} \bibnamefont{Chuu}}, \bibnamefont{and}
  \bibinfo{author}{\bibfnamefont{M.-Y.} \bibnamefont{Li}},
  \bibinfo{journal}{Sci. Adv.} \textbf{\bibinfo{volume}{3}},
  \bibinfo{pages}{e1601459} (\bibinfo{year}{2017}).

\bibitem[{\citenamefont{Liu et~al.}(2015{\natexlab{a}})\citenamefont{Liu, Jiao,
  Xie, Yang, Chen, Ho, Gao, Jia, Cui, and Xie}}]{Liu2015}
\bibinfo{author}{\bibfnamefont{H.~J.} \bibnamefont{Liu}},
  \bibinfo{author}{\bibfnamefont{L.}~\bibnamefont{Jiao}},
  \bibinfo{author}{\bibfnamefont{L.}~\bibnamefont{Xie}},
  \bibinfo{author}{\bibfnamefont{F.}~\bibnamefont{Yang}},
  \bibinfo{author}{\bibfnamefont{J.~L.} \bibnamefont{Chen}},
  \bibinfo{author}{\bibfnamefont{W.~K.} \bibnamefont{Ho}},
  \bibinfo{author}{\bibfnamefont{C.~L.} \bibnamefont{Gao}},
  \bibinfo{author}{\bibfnamefont{J.~F.} \bibnamefont{Jia}},
  \bibinfo{author}{\bibfnamefont{X.~D.} \bibnamefont{Cui}}, \bibnamefont{and}
  \bibinfo{author}{\bibfnamefont{M.~H.} \bibnamefont{Xie}},
  \bibinfo{journal}{2D Mater.} \textbf{\bibinfo{volume}{2}},
  \bibinfo{pages}{034004} (\bibinfo{year}{2015}{\natexlab{a}}).

\bibitem[{\citenamefont{Liu et~al.}(2015{\natexlab{b}})\citenamefont{Liu,
  Zheng, Yang, Jiao, Chen, Ho, Gao, Jia, and Xie}}]{Liu2015a}
\bibinfo{author}{\bibfnamefont{H.}~\bibnamefont{Liu}},
  \bibinfo{author}{\bibfnamefont{H.}~\bibnamefont{Zheng}},
  \bibinfo{author}{\bibfnamefont{F.}~\bibnamefont{Yang}},
  \bibinfo{author}{\bibfnamefont{L.}~\bibnamefont{Jiao}},
  \bibinfo{author}{\bibfnamefont{J.}~\bibnamefont{Chen}},
  \bibinfo{author}{\bibfnamefont{W.}~\bibnamefont{Ho}},
  \bibinfo{author}{\bibfnamefont{C.}~\bibnamefont{Gao}},
  \bibinfo{author}{\bibfnamefont{J.}~\bibnamefont{Jia}}, \bibnamefont{and}
  \bibinfo{author}{\bibfnamefont{M.}~\bibnamefont{Xie}}, \bibinfo{journal}{ACS
  Nano} \textbf{\bibinfo{volume}{9}}, \bibinfo{pages}{6619}
  (\bibinfo{year}{2015}{\natexlab{b}}).

\bibitem[{\citenamefont{P{\'{a}}sztor et~al.}(2017)\citenamefont{P{\'{a}}sztor,
  Scarfato, and Renner}}]{Pasztor2017}
\bibinfo{author}{\bibfnamefont{A.}~\bibnamefont{P{\'{a}}sztor}},
  \bibinfo{author}{\bibfnamefont{A.}~\bibnamefont{Scarfato}}, \bibnamefont{and}
  \bibinfo{author}{\bibfnamefont{C.}~\bibnamefont{Renner}},
  \bibinfo{journal}{Rev. Sci. Instrum.} \textbf{\bibinfo{volume}{88}},
  \bibinfo{pages}{3} (\bibinfo{year}{2017}).

\bibitem[{\citenamefont{Zhao et~al.}(2015)\citenamefont{Zhao, Wyrick, Natterer,
  Rodriguez-Nieva, Lewandowski, Watanabe, Taniguchi, Levitov, Zhitenev, and
  Stroscio}}]{Zhao2015}
\bibinfo{author}{\bibfnamefont{Y.}~\bibnamefont{Zhao}},
  \bibinfo{author}{\bibfnamefont{J.}~\bibnamefont{Wyrick}},
  \bibinfo{author}{\bibfnamefont{F.~D.} \bibnamefont{Natterer}},
  \bibinfo{author}{\bibfnamefont{J.~F.} \bibnamefont{Rodriguez-Nieva}},
  \bibinfo{author}{\bibfnamefont{C.}~\bibnamefont{Lewandowski}},
  \bibinfo{author}{\bibfnamefont{K.}~\bibnamefont{Watanabe}},
  \bibinfo{author}{\bibfnamefont{T.}~\bibnamefont{Taniguchi}},
  \bibinfo{author}{\bibfnamefont{L.~S.} \bibnamefont{Levitov}},
  \bibinfo{author}{\bibfnamefont{N.~B.} \bibnamefont{Zhitenev}},
  \bibnamefont{and} \bibinfo{author}{\bibfnamefont{J.~A.}
  \bibnamefont{Stroscio}}, \bibinfo{journal}{Science}
  \textbf{\bibinfo{volume}{348}}, \bibinfo{pages}{672} (\bibinfo{year}{2015}).

\bibitem[{\citenamefont{Qiao et~al.}(2014)\citenamefont{Qiao, Kong, Hu, Yang,
  and Ji}}]{Qiao2014}
\bibinfo{author}{\bibfnamefont{J.}~\bibnamefont{Qiao}},
  \bibinfo{author}{\bibfnamefont{X.}~\bibnamefont{Kong}},
  \bibinfo{author}{\bibfnamefont{Z.-X.} \bibnamefont{Hu}},
  \bibinfo{author}{\bibfnamefont{F.}~\bibnamefont{Yang}}, \bibnamefont{and}
  \bibinfo{author}{\bibfnamefont{W.}~\bibnamefont{Ji}}, \bibinfo{journal}{Nat.
  Commun.} \textbf{\bibinfo{volume}{5}}, \bibinfo{pages}{4475}
  (\bibinfo{year}{2014}).

\bibitem[{\citenamefont{Xia et~al.}(2019)\citenamefont{Xia, Wang, Hwang, Neto,
  and Yang}}]{Xia2019}
\bibinfo{author}{\bibfnamefont{F.}~\bibnamefont{Xia}},
  \bibinfo{author}{\bibfnamefont{H.}~\bibnamefont{Wang}},
  \bibinfo{author}{\bibfnamefont{J.~C.~M.} \bibnamefont{Hwang}},
  \bibinfo{author}{\bibfnamefont{A.~H.~C.} \bibnamefont{Neto}},
  \bibnamefont{and} \bibinfo{author}{\bibfnamefont{L.}~\bibnamefont{Yang}},
  \bibinfo{journal}{Nat. Rev. Phys.} \textbf{\bibinfo{volume}{1}},
  \bibinfo{pages}{306} (\bibinfo{year}{2019}).

\bibitem[{\citenamefont{Kim et~al.}(2018)\citenamefont{Kim, Seo, Lee, Ko, Kim,
  Jang, Ok, Lee, Jo, Kang et~al.}}]{Kim2018}
\bibinfo{author}{\bibfnamefont{K.}~\bibnamefont{Kim}},
  \bibinfo{author}{\bibfnamefont{J.}~\bibnamefont{Seo}},
  \bibinfo{author}{\bibfnamefont{E.}~\bibnamefont{Lee}},
  \bibinfo{author}{\bibfnamefont{K.-T.} \bibnamefont{Ko}},
  \bibinfo{author}{\bibfnamefont{B.~S.} \bibnamefont{Kim}},
  \bibinfo{author}{\bibfnamefont{B.~G.} \bibnamefont{Jang}},
  \bibinfo{author}{\bibfnamefont{J.~M.} \bibnamefont{Ok}},
  \bibinfo{author}{\bibfnamefont{J.}~\bibnamefont{Lee}},
  \bibinfo{author}{\bibfnamefont{Y.~J.} \bibnamefont{Jo}},
  \bibinfo{author}{\bibfnamefont{W.}~\bibnamefont{Kang}}, \bibnamefont{et~al.},
  \bibinfo{journal}{Nat. Mater.} \textbf{\bibinfo{volume}{17}},
  \bibinfo{pages}{794} (\bibinfo{year}{2018}).

\bibitem[{\citenamefont{Kim et~al.}(2019)\citenamefont{Kim, Park, Lee, Yoon,
  Joo, Kim, Min, Park, Kim, Moon et~al.}}]{Kim2019}
\bibinfo{author}{\bibfnamefont{D.}~\bibnamefont{Kim}},
  \bibinfo{author}{\bibfnamefont{S.}~\bibnamefont{Park}},
  \bibinfo{author}{\bibfnamefont{J.}~\bibnamefont{Lee}},
  \bibinfo{author}{\bibfnamefont{J.}~\bibnamefont{Yoon}},
  \bibinfo{author}{\bibfnamefont{S.}~\bibnamefont{Joo}},
  \bibinfo{author}{\bibfnamefont{T.}~\bibnamefont{Kim}},
  \bibinfo{author}{\bibfnamefont{K.-j.} \bibnamefont{Min}},
  \bibinfo{author}{\bibfnamefont{S.-Y.} \bibnamefont{Park}},
  \bibinfo{author}{\bibfnamefont{C.}~\bibnamefont{Kim}},
  \bibinfo{author}{\bibfnamefont{K.-W.} \bibnamefont{Moon}},
  \bibnamefont{et~al.}, \bibinfo{journal}{Nanotechnology}
  \textbf{\bibinfo{volume}{30}}, \bibinfo{pages}{245701}
  (\bibinfo{year}{2019}).

\bibitem[{\citenamefont{May et~al.}(2016)\citenamefont{May, Calder, Cantoni,
  Cao, and McGuire}}]{May2016}
\bibinfo{author}{\bibfnamefont{A.~F.} \bibnamefont{May}},
  \bibinfo{author}{\bibfnamefont{S.}~\bibnamefont{Calder}},
  \bibinfo{author}{\bibfnamefont{C.}~\bibnamefont{Cantoni}},
  \bibinfo{author}{\bibfnamefont{H.}~\bibnamefont{Cao}}, \bibnamefont{and}
  \bibinfo{author}{\bibfnamefont{M.~A.} \bibnamefont{McGuire}},
  \bibinfo{journal}{Phys. Rev. B} \textbf{\bibinfo{volume}{93}},
  \bibinfo{pages}{014411} (\bibinfo{year}{2016}).

\bibitem[{\citenamefont{Fei et~al.}(2018)\citenamefont{Fei, Huang, Malinowski,
  Wang, Song, Sanchez, Yao, Xiao, Zhu, May et~al.}}]{Fei2018}
\bibinfo{author}{\bibfnamefont{Z.}~\bibnamefont{Fei}},
  \bibinfo{author}{\bibfnamefont{B.}~\bibnamefont{Huang}},
  \bibinfo{author}{\bibfnamefont{P.}~\bibnamefont{Malinowski}},
  \bibinfo{author}{\bibfnamefont{W.}~\bibnamefont{Wang}},
  \bibinfo{author}{\bibfnamefont{T.}~\bibnamefont{Song}},
  \bibinfo{author}{\bibfnamefont{J.}~\bibnamefont{Sanchez}},
  \bibinfo{author}{\bibfnamefont{W.}~\bibnamefont{Yao}},
  \bibinfo{author}{\bibfnamefont{D.}~\bibnamefont{Xiao}},
  \bibinfo{author}{\bibfnamefont{X.}~\bibnamefont{Zhu}},
  \bibinfo{author}{\bibfnamefont{A.~F.} \bibnamefont{May}},
  \bibnamefont{et~al.}, \bibinfo{journal}{Nat. Mater.}
  \textbf{\bibinfo{volume}{17}}, \bibinfo{pages}{778} (\bibinfo{year}{2018}).

\bibitem[{\citenamefont{Deng et~al.}(2018)\citenamefont{Deng, Yu, Song, Zhang,
  Wang, Sun, Yi, Wu, Wu, Zhu et~al.}}]{Deng2018}
\bibinfo{author}{\bibfnamefont{Y.}~\bibnamefont{Deng}},
  \bibinfo{author}{\bibfnamefont{Y.}~\bibnamefont{Yu}},
  \bibinfo{author}{\bibfnamefont{Y.}~\bibnamefont{Song}},
  \bibinfo{author}{\bibfnamefont{J.}~\bibnamefont{Zhang}},
  \bibinfo{author}{\bibfnamefont{N.~Z.} \bibnamefont{Wang}},
  \bibinfo{author}{\bibfnamefont{Z.}~\bibnamefont{Sun}},
  \bibinfo{author}{\bibfnamefont{Y.}~\bibnamefont{Yi}},
  \bibinfo{author}{\bibfnamefont{Y.~Z.} \bibnamefont{Wu}},
  \bibinfo{author}{\bibfnamefont{S.}~\bibnamefont{Wu}},
  \bibinfo{author}{\bibfnamefont{J.}~\bibnamefont{Zhu}}, \bibnamefont{et~al.},
  \bibinfo{journal}{Nature} \textbf{\bibinfo{volume}{563}}, \bibinfo{pages}{94}
  (\bibinfo{year}{2018}).

\bibitem[{not()}]{note}
\bibinfo{note}{For quick test purpose, the stacked G/hBN/Gr device was prepared
  outside a glove box since all materials are inert in air. In principle, we
  can prepare a stack of air-sensitive materials in the same way inside a glove
  box}.

\end{thebibliography}

\end{document}